# Compression benchmarking of holotomography data using the OME-Zarr storage format

Dohyeon Lee[1,2], Juyeon Park[1,2], Juheon Lee[1,2], Chungha Lee[1,2], and YongKeun Park[1,2,3*]

[1]*Department of Physics, Korea Advanced Institute of Science and Technology (KAIST), Daejeon, Republic of Korea*

[2]*KAIST Institute for Health Science and Technology, KAIST, Daejeon, Republic of Korea*

[3]*Tomocube Inc., Daejeon, Republic of Korea*

[*]*yk.park@kaist.ac.kr*

**Abstract**

Holotomography (HT) is a label-free, three-dimensional imaging technique that captures refractive index distributions of biological samples at sub-micron resolution. As modern HT systems enable high-throughput and large-scale acquisition, they produce terabyte-scale datasets that require efficient data management. This study presents a systematic benchmarking of data compression strategies for HT data stored in the OME-Zarr format, a cloud-compatible, chunked data structure suitable for scalable imaging workflows. Using representative datasets—including embryo, tissue, and birefringent tissue volumes—we evaluated combinations of preprocessing filters and 25 compression configurations across multiple compression levels. Performance was assessed in terms of compression ratio, bandwidth, and decompression speed. A throughput-based evaluation metric was introduced to simulate real-world conditions under varying network constraints, supporting optimal compressor selection based on system bandwidth. The results offer practical guidance for storage and transmission of large HT datasets and serve as a reference for implementing scalable, FAIR-aligned imaging workflows in cloud and high-performance computing environments.

**Introduction**

Holotomography (HT), also known as three-dimensional (3D) quantitative phase imaging, has emerged as a powerful, label-free, high-resolution imaging modality. Holotomography reconstructs the 3D refractive index (RI) distribution of unlabeled live biological samples from multiple two-dimensional light-field measurements, analogous to X-ray computed tomography[1-6]. Its quantitative, tomographic, and noninvasive imaging capabilities have enabled broad applications across cell biology[7,8], biophysics[9,10], neuroscience[11], microbiology[12], immunology[13], regenerative medicine[14-16], and histopathology[17].

Recent advances in optical systems and reconstruction algorithms have significantly improved the speed and scalability of HT, making it suitable for high-throughput applications. State-of-the-art HT systems can acquire a 3D image volume of 200 μm × 200 μm × 150 μm at spatial resolutions of 150 nm × 150 nm × 800 nm within 1 second. Lateral stitching techniques now enable imaging of centimeter-scale tissue sections[17], while robust long-term imaging allows time-lapse acquisition spanning several weeks[18]. Moreover, the recent development of birefringence-sensitive HT systems enables the reconstruction of 3×3 tensor RI representations at each voxel, facilitating a deeper understanding of molecular alignment and anisotropic structures in biological samples[19-23].

However, the ability of HT to generate extremely large datasets over extended time periods introduces substantial challenges in data storage, transmission, and real-time accessibility (Fig. 1). Traditional image formats such as TIFF, although structured, are limited in scalability, metadata management, and cloud compatibility due to their monolithic file structures and lack of efficient chunking or encoding strategies (Fig. 1A). HDF5, while supporting multi-dimensional arrays, suffers from limited support for real-time or remote access, as entire datasets typically need to be downloaded prior to analysis[24,25].

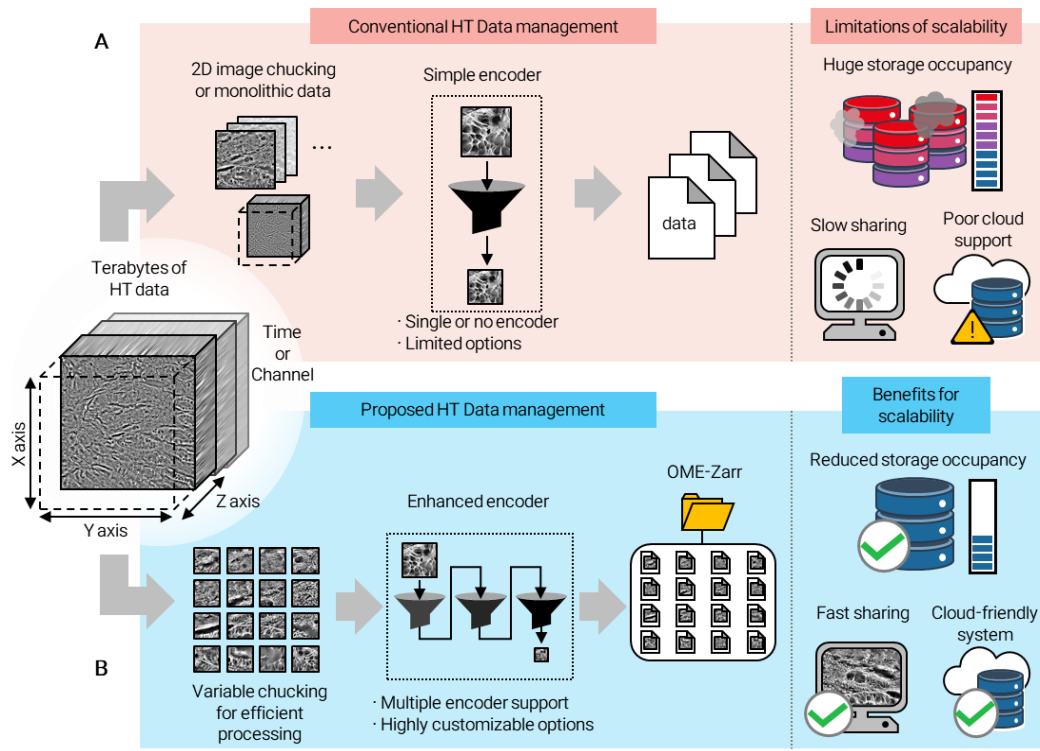

**Fig. 1 | Comparison of conventional and proposed data management strategies for holotomography (HT) imaging.** A, Conventional methods use simple or no encoding with limited chunking, resulting in large storage needs and poor scalability. B, The proposed approach employs variable chunking, customizable encoders, and OME-Zarr format, enabling efficient storage, fast sharing, and cloud compatibility.

To overcome these limitations, Zarr, a cloud-native, chunked, and compressed data storage format, has gained attention as a promising alternative for managing terabyte-scale imaging data[25,26]. Zarr divides data into user-defined chunks that can be processed and stored independently, enabling parallel read/write operations, distributed computing, and seamless integration with cloud infrastructure. Users can define custom chunking strategies for their specific needs and optimize performance. Before converting a chunk of data into a file, it can be processed through various encoding options, including preprocessing filters and compression algorithms, allowing for efficient storage utilization and faster data sharing. The OME-Zarr specification extends this format for multi-dimensional bioimaging by incorporating standardized metadata for dimensionality, resolution, and channel information, thereby supporting complex imaging workflows in scalable and FAIR(Findable, Accessible, Interoperable, and Reusable)-compliant environments[27].

Despite the advantages of Zarr-based storage, the selection of an optimal compression strategy remains nontrivial. Efficient encoding not only reduces storage footprint but also directly impacts data access speed—an increasingly critical factor in AI-driven analysis, real-time processing, and cloud-native applications[28]. Although prior studies have reported compression benchmarks in other imaging contexts[29-31], a comprehensive evaluation tailored to HT imaging has not been performed.

In this study, we present a systematic benchmarking analysis of compression strategies for HT data stored in OME-Zarr format (Fig. 1B). We assess the effects of preprocessing filters (bit-depth rounding and byte shuffling) and compare the performance of widely used compression algorithms, including gzip, LZMA, bzip2, zlib, LZ4, zstd, and Blosc-based methods[32-38]. Our evaluation spans diverse datasets, including both conventional and birefringent HT volumes, across conditions of varying spatial density and signal strength. We measure and compare compression ratio, compression bandwidth, and decompression bandwidth under realistic usage scenarios, including simulated cloud-based access.

To provide practical guidance for diverse computational environments, we introduce a throughput-based evaluation metric that integrates network transfer bandwidth with compression and decompression speeds. This metric enables a context-aware comparison of compression strategies, helping users select optimal configurations tailored to specific storage and data transfer conditions.

Our findings offer key insights for scalable, high-performance HT data management, supporting efficient storage, rapid access, and smooth integration with modern computational pipelines. This work provides a foundational resource for HT imaging workflows and contributes to broader efforts in building FAIR-aligned data infrastructures for large-scale biomedical imaging.

**Results**

**Evaluation of compression ratio**

To evaluate the effects of various filtering and compression strategies on HT data, we analyzed the overall compression ratio across different sample types, preprocessing filters, and compression algorithms (Fig. 2). The left panel of each plot in Fig. 2 presents the average compression ratios for four filtering configurations: no filter, byte shuffle, bit rounding, and a combination of bit rounding and byte shuffle. The right panels display the full distribution of compression outcomes, providing a more detailed view of algorithm-specific performance under different conditions.

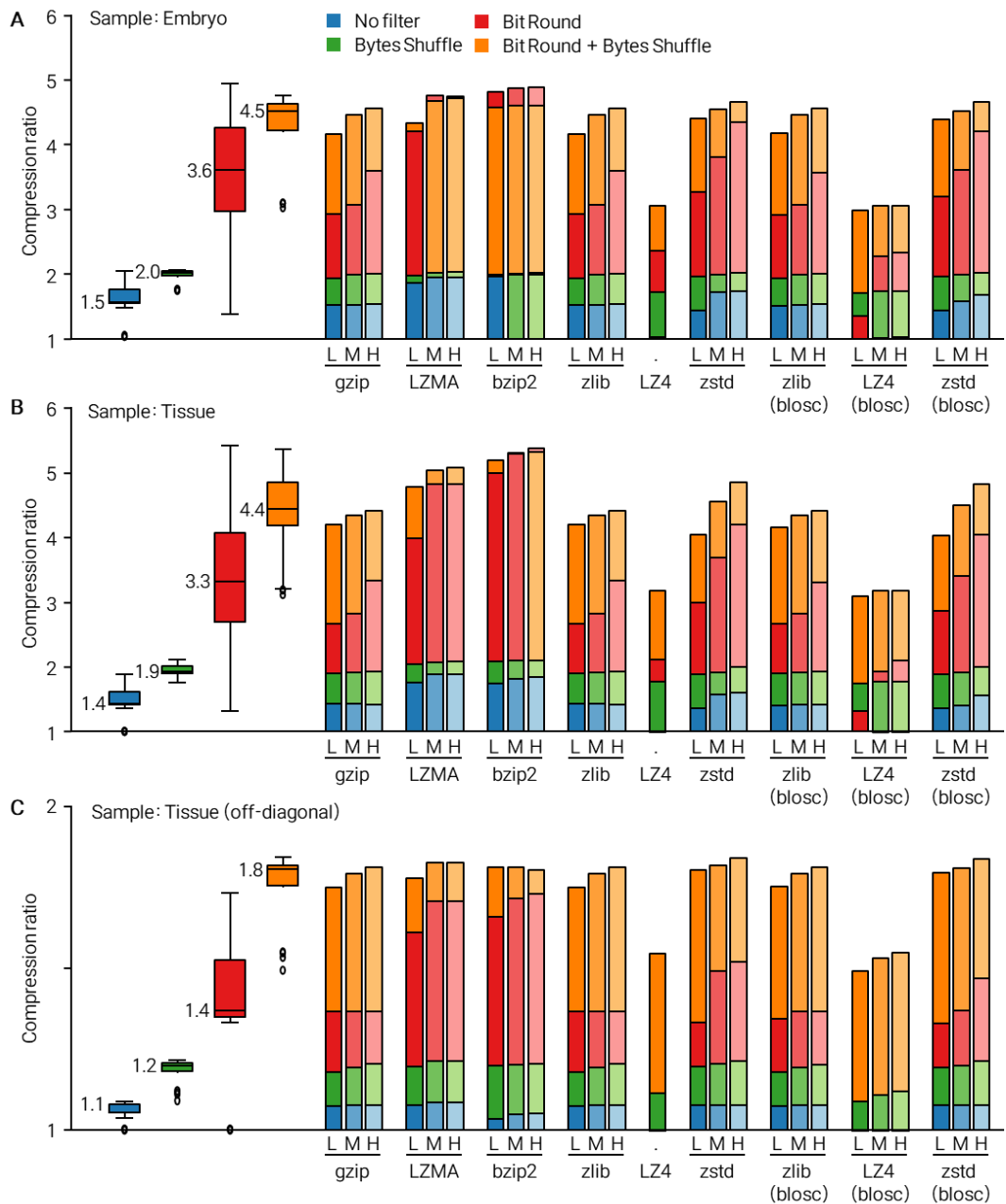

**Fig. 2 | Compression ratio comparison across datasets and filtering methods.** A, Embryo data, B, Tissue data, C, Tissue (off-diagonal) data. Left panels show average compression ratios for each filter: no filter (blue), byte shuffle (green), bit round (red), and combined bit round + byte shuffle (orange). Stacked bars indicate results by compression algorithm and level: low (L), medium (M), and high (H).

Across all datasets (embryo, tissue, and tissue off-diagonal data), filtering improved compression efficiency by increasing data regularity and redundancy. In particular, bit rounding consistently yielded higher compression gains than byte shuffling alone. This is because bit rounding eliminates scientifically insignificant bits by setting them to zero, introducing highly redundant zero-value regions without compromising meaningful information. Since many scientific measurements are limited by the precision of the instrument, reducing excess precision in this way is often justifiable and effective.

In contrast, byte shuffling enhances compressibility by reorganizing the byte order within data chunks, bringing together similar byte positions and making patterns more apparent to compression algorithms.

This technique is most effective when adjacent voxel values are similar or change gradually—conditions commonly found in HT images due to the inherently slow spatial variation in biological samples, which are also leveraged during HT reconstruction[39].

Thus, while byte shuffling preserves all information (lossless), bit rounding introduces controlled loss (lossy) for improved compression. The optimal filtering strategy depends on whether strict data fidelity is required. Table 1 summarizes the recommended filtering combinations based on the desired compression type.

|                        | Lossless      | Lossy                       |
| ---------------------- | ------------- | --------------------------- |
| Best filter combination | Bytes Shuffle | Bit Round + Bytes Shuffle   |

Table 1 | Recommended filter combinations for lossless and lossy compression of holotomographic data.

Across most algorithms, the compression ratio followed a consistent trend: no filtering < byte shuffle < bit rounding < bit rounding + byte shuffle. However, certain exceptions were observed for specific algorithms such as LZMA, bzip2, and LZ4, due to their distinct internal compression mechanisms.

For instance, LZMA improves upon gzip's dictionary-matching and entropy coding approach by identifying long-range redundancy. Byte shuffling can disrupt these long-range patterns by altering the byte sequence, reducing its effectiveness[33]. Similarly, bzip2 applies the Burrows–Wheeler transform prior to compression, which already reorders data to enhance compressibility[34,40]. The added byte shuffling may conflict with this transformation's internal structure, leading to suboptimal performance gains.

In the case of LZ4, which is optimized for speed and short-range pattern detection, bit rounding did not substantially improve the compression of off-diagonal tissue data. Because off-diagonal values are typically small and noise-sensitive, the local patterns that LZ4 relies on remain obscured. However, LZ4 showed marked improvement when combined with byte shuffling, which better exposed short-range repetitions within compressed chunks[36]. These algorithm-specific behaviors highlight the nontrivial interaction between preprocessing filters and compression backends.

Despite the general consistency in filtering effectiveness across sample types, differences in compression ratio across datasets reveal important insights into how sample-specific characteristics influence data compressibility. Notably, while embryo and tissue datasets showed comparable average compression ratios, the off-diagonal tensor components from birefringent tissue data exhibited a 55% lower compression ratio on average. This reduction likely stems from higher noise susceptibility in these off-diagonal terms, as their values are typically smaller and more variable. Compression algorithms, which rely on pattern detection, are inherently less effective in the presence of noise, which increases entropy and reduces redundancy.

These findings suggest that further improvements in compressibility for noise-prone datasets could be achieved through additional preprocessing, such as denoising or more aggressive precision reduction beyond standard bit rounding. Future work may explore these options to improve storage efficiency without compromising data utility in downstream analyses.

**Evaluation of Compression and Decompression Bandwidth**

Beyond compression ratio, the bandwidth of compression and decompression plays a pivotal role in determining the practical efficiency of a compression strategy. While higher compression ratios enhance storage efficiency, suboptimal speed performance can introduce significant computational bottlenecks, especially in real-time processing or latency-sensitive workflows. The utility of HT data lies not only in how compactly it can be stored, but also in how quickly it can be accessed and processed.

For instance, decompression speed directly impacts the responsiveness of downstream tasks such as quantitative analysis, 3D visualization, and AI inference. In cloud-based HT workflows—where reconstruction and analysis are performed on remote servers—the combination of compression and decompression bandwidth dictates how quickly results can be transmitted to and accessed by end users. These scenarios illustrate the need to optimize both bandwidth and compression efficiency to ensure a seamless user experience and effective integration into high-throughput, distributed imaging pipelines.

To isolate bandwidth performance from compression ratio effects, we conducted speed benchmarks using byte shuffling only, as this filtering strategy showed the most stable and consistent compression ratios across datasets (Fig. 3). All benchmarks were executed in a simulated parallel compression environment to reflect realistic usage of chunked storage in OME-Zarr, where parallelized processing is often leveraged. A comprehensive benchmark including all filtering combinations is available in Supplementary Table 1.

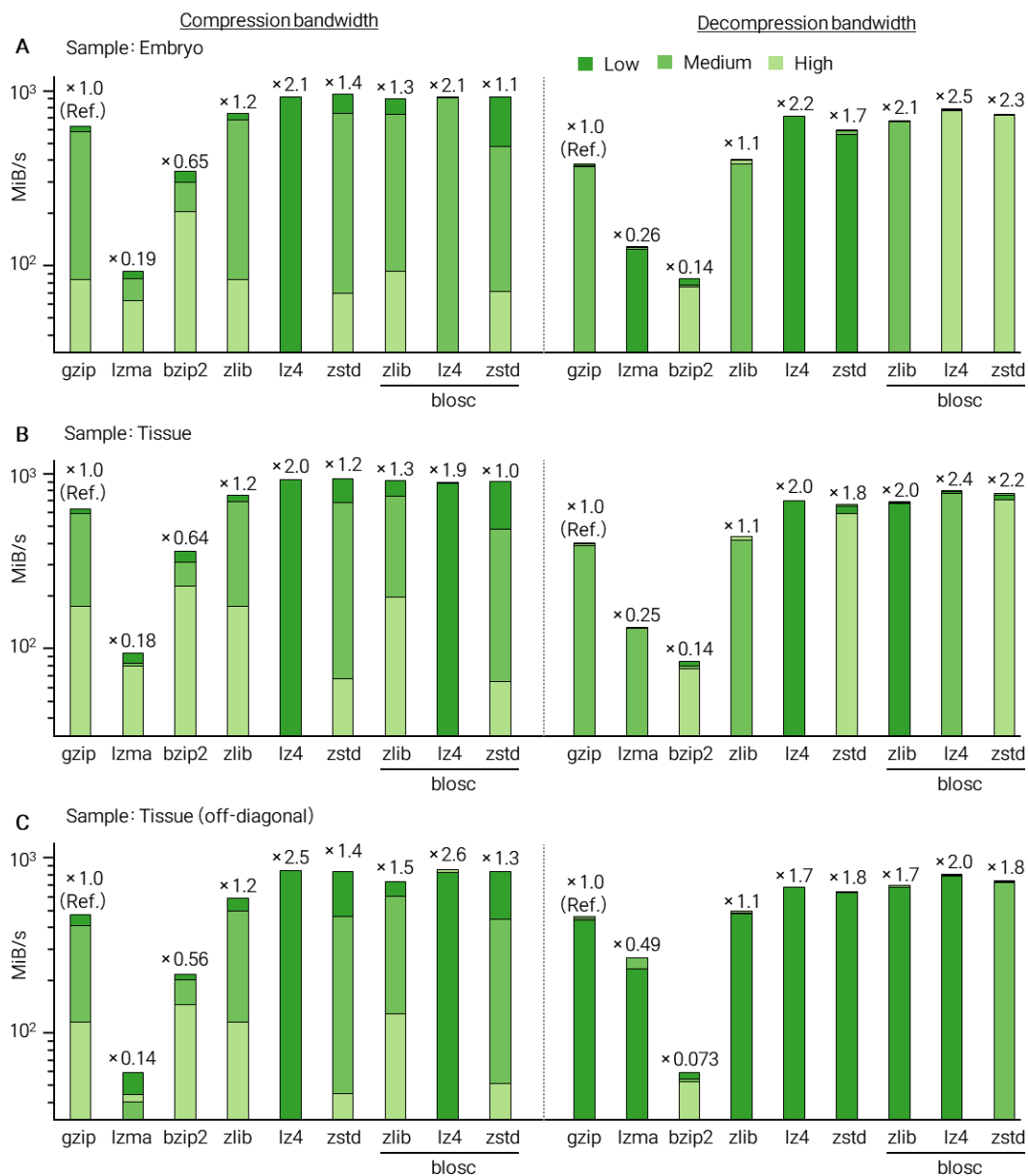

**Fig. 3 | Compression and decompression speed benchmarks with byte shuffle enabled.** A, Embryo data, B, Tissue data, C, Tissue (off-diagonal) data. Left panels show compression bandwidths, and right panels show decompression bandwidths, measured in mebibytes per second (MiB/s). Values above each bar indicate relative speed compared to gzip, averaged across compression levels.

The compression bandwidth results reveal that modern, speed-optimized algorithms such as LZ4, Zstandard (zstd), and Blosc-based variants exhibit superior performance, particularly at low compression levels. These algorithms achieve significantly higher throughput compared to traditional methods such as gzip, bzip2, or LZMA. However, their performance advantage diminishes as compression level increases. For example, at high compression levels, zstd's compression bandwidth becomes comparable to, or even lower than, that of conventional methods. This suggests a trade-off: speed-optimized algorithms incur increasing computational overhead at higher compression ratios, potentially offsetting their usual performance benefits. Therefore, both the choice of compression algorithm and its configuration level must be considered in tandem to optimize processing speed.

In contrast, decompression benchmarks showed consistent superiority of speed-optimized algorithms across all datasets and compression levels. Regardless of configuration, LZ4, zstd, and Blosc-based methods demonstrated significantly higher decompression bandwidths compared to classic algorithms. This indicates that modern compressors are better engineered for fast data restoration—an especially valuable trait in scenarios where rapid access to compressed data is critical.

**Definition of a Throughput-Based Metric for HT Compression Evaluation**

Since no single benchmark parameter can independently determine the optimal compression strategy, we introduce a unified, throughput-based evaluation metric designed to reflect the practical demands of future HT data workflows. Modern data management in HT extends beyond simple storage and sharing, increasingly functioning as a platform for reconstruction, visualization, and multidimensional analysis. In this evolving landscape, it is anticipated that users will offload computationally intensive tasks—such as 3D reconstruction and polarization analysis—to high-performance cloud or server-based platforms[41,42].

Our proposed metric models a typical cloud-based HT reconstruction and analysis scenario, leveraging the chunk-level parallelism supported by the OME-Zarr format. Unlike traditional file-based workflows, where compression, transfer, and decompression times are additive, chunked parallel workflows are constrained by the slowest of these operations. Thus, our metric focuses on identifying the bottleneck step and quantifying the system's maximum sustainable throughput, offering a realistic measure of overall performance in distributed HT imaging pipelines.

We define the effective bandwidth (EB) as the product of the compression ratio (CR) and the available transfer bandwidth (TB), representing the logical throughput gain achieved via compression:

$$EB = CR \times TB.$$

Here, CR is computed as the average compression ratio obtained using either byte shuffling alone or a combination of byte shuffling and bit rounding, thus capturing both lossless and lossy compression scenarios. This average allows for an inclusive evaluation that accounts for precision-sensitive and precision-tolerant applications.

Next, we define the maximum achievable bandwidth ($B_{max}$) as the minimum value among the effective bandwidth, compression bandwidth (CB), and decompression bandwidth (DB):

$$B_{max} = \min(EB, CB, DB) = \min(CR \times TB, CB, DB).$$

This formulation reflects the fact that system throughput is inherently limited by the slowest stage in the compression–transfer–decompression pipeline. Therefore, $B_{max}$ serves as a unified indicator of end-to-end performance and provides a data-driven basis for selecting compression strategies suited to real-world HT data management scenarios, particularly in cloud-based and remote processing environments.

This metric provides a systematic basis for selecting the optimal compression method under varying data transfer constraints, making it particularly relevant for high-throughput HT imaging applications in cloud-based environments.

**Application of the Metric Across Varying Network Bandwidths**

To illustrate the practical implications of $B_{max}$, we computed and ranked compression algorithms under varying network transfer bandwidths (Fig. 4). The results highlight the dynamic nature of optimal compressor selection based on bandwidth availability:

- Under low transfer bandwidth conditions, compression algorithms with higher compression ratios—such as bzip2 and LZMA—perform better, as the reduced data volume offsets their slower processing speeds.

- At high bandwidth levels (e.g., in high-speed server clusters with dedicated interconnects), fast compression and decompression algorithms like Blosc-LZ4 or LZ4 (at low compression levels) are favored. In these cases, data transfer is no longer the limiting factor, so maximizing processing speed becomes the priority.

- At mid-range bandwidths (e.g., 1–10 Gbps, typical of modern Ethernet infrastructure standardized in 1999 and 2006[43,44]), balanced strategies are preferred. Compression methods that offer consistent, moderate performance across all metrics are most advantageous in these commonly encountered environments.

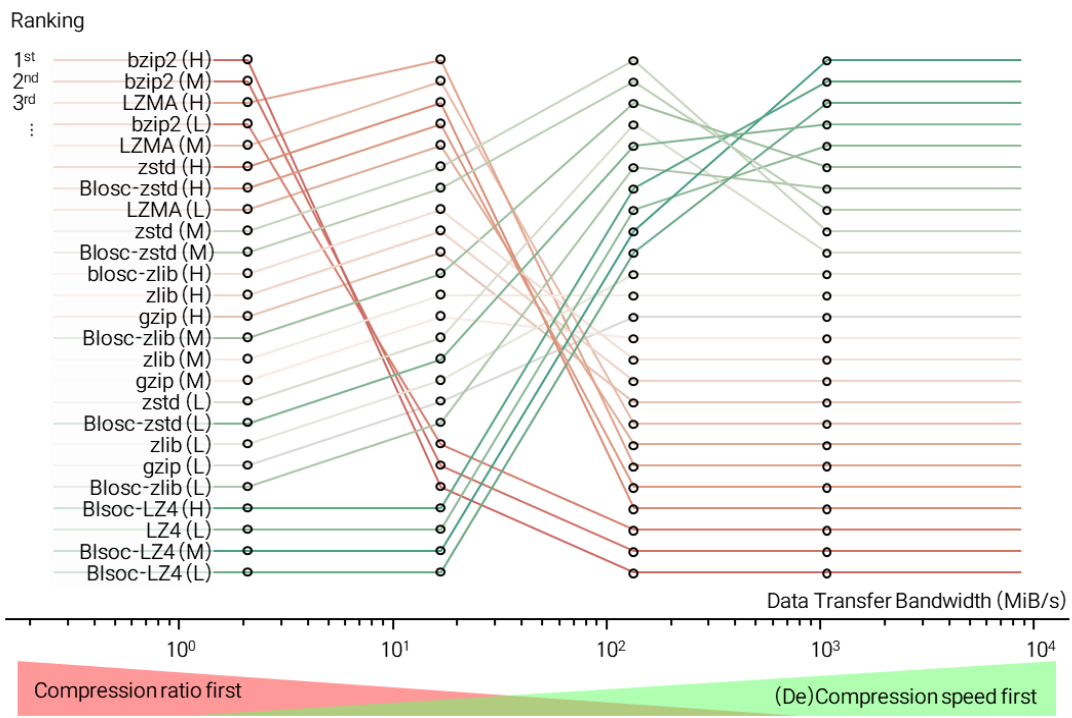

**Fig. 4 | Ranking of compression methods based on a throughput-optimized metric under varying data transfer bandwidths.** Compression methods are ranked from left (compression ratio prioritized) to right ((de)compression speed prioritized), reflecting optimal choices depending on network or I/O constraints.

Compression algorithms are ordered from left to right based on shifting emphasis from compression ratio (low-bandwidth scenarios) to (de)compression speed (high-bandwidth scenarios), reflecting their optimal applicability under different network constraints.

For systems subject to fluctuating network conditions or requiring broad interoperability, it is essential to select a compressor that performs reliably across diverse bandwidth scenarios. To address this, we evaluated the worst-case performance of each compression algorithm across multiple transfer bandwidths. The top-performing "all-rounder" compression methods are summarized in Table 2.

| Ranking | 1st | 2nd | 3rd |
| --- | --- | --- | --- |
| All-rounder compression option | zstd (M) | blosc-zstd (M) | blosc-zlib (M) |

Table 2 | All-rounder compression options based on the throughput-based metric.

These compression options strike an effective balance between high compression ratios and fast processing speeds, making them suitable for a wide array of applications including cloud-based HT reconstruction, streaming visualization, and AI-powered scientific computing.

**Discussion**

In this study, we systematically evaluated data compression strategies for HT imaging within the OME-Zarr framework, aiming to optimize storage efficiency and facilitate scalable data sharing. As HT imaging continues to produce increasingly large and high-resolution datasets, effective compression has become essential—not only for minimizing storage footprints and data transfer costs, but also for enabling seamless accessibility in collaborative, cloud-based research environments.

Given the chunked and cloud-native nature of OME-Zarr, we investigated how different combinations of preprocessing filters and compression algorithms influence both storage and computational performance in realistic HT workflows. Our benchmark results demonstrate that the optimal choice of compression strategy depends heavily on the available data transfer bandwidth. In low-bandwidth scenarios, algorithms with higher compression ratios (e.g., bzip2, LZMA) reduce transmission time despite slower processing speeds. Conversely, in high-bandwidth settings, fast compression methods (e.g., Blosc-LZ4, LZ4) are more effective due to reduced computational bottlenecks. For bandwidth conditions typical of modern Ethernet environments (1–10 Gbps), balanced options such as zstd (Medium), Blosc-zstd (Medium), and Blosc-zlib (Medium) provide robust performance across metrics, making them suitable for general-purpose use.

Future work should expand upon this benchmarking by exploring additional compression methods and filtering strategies. Multi-dimensional and lossy scientific compressors such as Blosc2, Pcodec, and ZFP—designed specifically for numerical data—could offer greater storage savings while preserving acceptable precision levels[38,45,46]. Another important direction is optimizing chunk size and shape, as these parameters significantly impact read/write performance and access latency in parallel or cloud-based data processing workflows[47].

Beyond performance optimization, our findings highlight the broader advantages of adopting the Zarr and OME-Zarr ecosystem over legacy formats such as TIFF and HDF5. These next-generation formats offer a simplified design, excellent support for distributed access, and compatibility with cloud infrastructure. Despite being relatively new, Zarr has already been implemented across multiple programming languages and platforms, and its development continues through active contributions from a wide community of researchers, developers, and industry stakeholders.

The HT imaging community is well-positioned to play a proactive role in shaping the future of Zarr-based data management. By contributing domain-specific use cases, feature requests, and performance insights, HT researchers can help tailor the evolution of OME-Zarr to meet the unique demands of high-throughput, multidimensional imaging.

In conclusion, this study provides a comprehensive baseline for compression strategy selection within the OME-Zarr framework and supports the ongoing development of FAIR, high-performance data infrastructure for holotomography. Through collaborative and open innovation, the community can continue to enhance data sharing, analysis, and archiving practices across the biomedical imaging field.

## Methods

### Benchmarking process overview

To systematically assess compression performance for HT data, we designed a benchmarking workflow based on chunked OME-Zarr datasets (Fig. 5a). All datasets were stored in single-precision floating-point format, and chunk size was fixed at 32 × 256 × 256 along the z, y, and x dimensions, respectively. This configuration ensured a uniform data structure across all test cases, with each uncompressed chunk occupying exactly 8 MiB of memory.

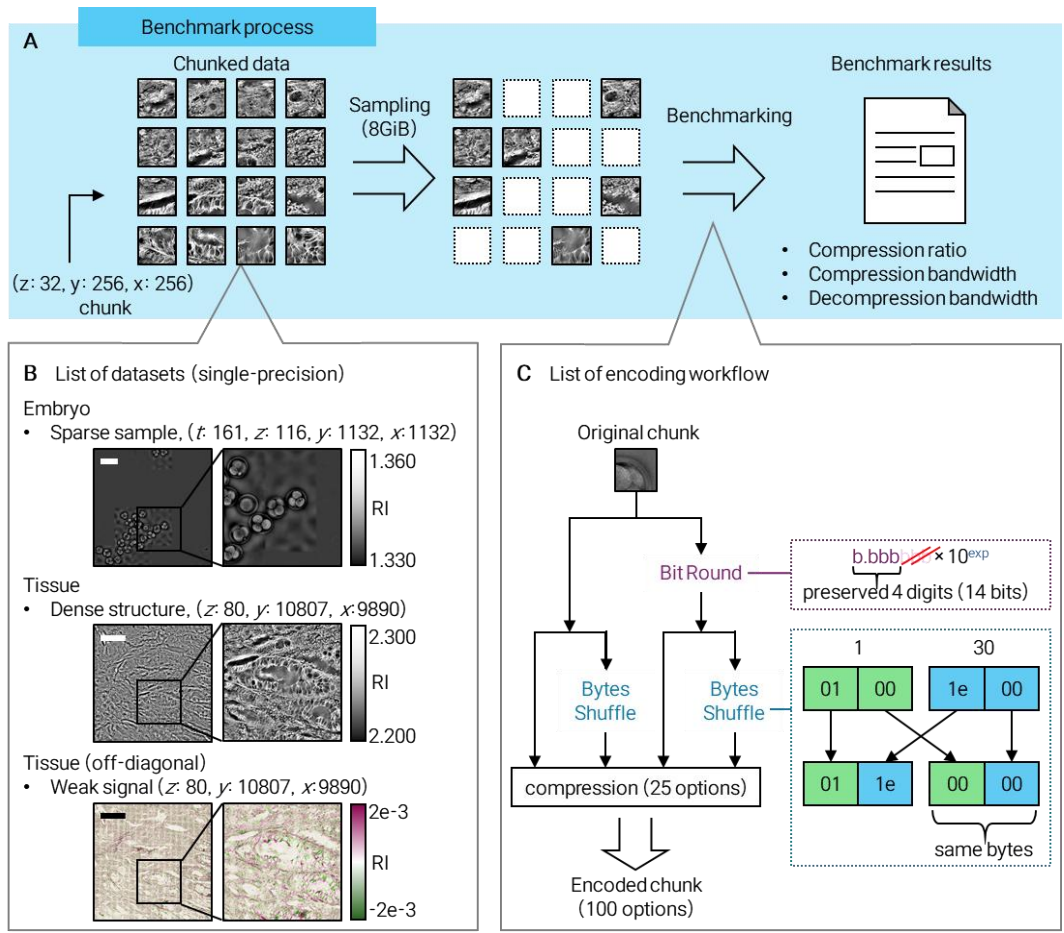

**Fig. 5 | Benchmarking workflow for HT data compression.** A, Overview of the benchmarking process with fixed chunk shape and sampling size (8 GiB). B, Representative datasets used in the benchmark, all stored as single-precision float. C, Encoding workflow illustrating applied filters, including bit rounding and byte shuffling, followed by compression with 25 different options.

Due to the large volume of the original HT datasets, benchmarking was performed on a sampled subset totaling 8 GiB of data, corresponding to 1024 randomly selected chunks. This sampling strategy allowed us to perform compression tests efficiently while preserving data diversity representative of real-world scenarios. For each compression configuration, we evaluated three core performance metrics: compression ratio, compression bandwidth, and decompression bandwidth, enabling a comprehensive analysis of both storage efficiency and computational throughput.

Benchmarks were executed in a multi-threaded environment to reflect common usage patterns in OME-Zarr pipelines, which typically leverage parallel I/O and chunk-level operations. Consequently, the reported bandwidth metrics represent peak throughput under full CPU utilization. To isolate computational performance from potential disk I/O bottlenecks, all compression and decompression tasks were performed entirely in memory. Input data chunks were preloaded into RAM, compressed data was written to separate memory buffers, and decompression was conducted by reading directly from those buffers. This design ensured that the measurements reflect true processing performance, unaffected by storage latency.

All tests were conducted on a high-performance workstation equipped with an AMD Ryzen Threadripper 2950X processor (32 threads) and 128 GB of DDR4 RAM, providing ample parallel computing capacity for multi-threaded compression workloads. This setup ensured reproducibility and stability during benchmark execution across all tested configurations.

**Dataset preparation**

To evaluate compression performance under realistic and heterogeneous imaging conditions, the benchmarking was conducted using holotomographic (HT) datasets that include intrinsic and artificially introduced noise. Three representative types of datasets were selected: Embryo, Tissue, and Tissue (off-diagonal). These datasets capture a range of biological sample properties, varying in spatial density, refractive index contrast, and signal-to-noise characteristics (Fig. 5b).

The Embryo dataset was acquired using Tomocube's HT-X1™ imaging system, optimized for live embryo visualization[16]. All experimental procedures were conducted in accordance with the Institutional Animal Care and Use Committee (IACUC) guidelines (KA2023-005-v4). The raw HT data was stored in single-precision floating-point format, with meaningful values preserved up to four decimal places. To simulate raw acquisition noise that is typically present in unprocessed HT data, uniform random noise in the range of $-5 \times 10^{-5}$ to $+5 \times 10^{-5}$ was added to each voxel prior to compression benchmarking. This approach ensures a more realistic evaluation of compression performance in practical imaging scenarios.

The Tissue dataset was obtained using a custom-built HT microscope designed for high-resolution, birefringence-sensitive imaging[19,48]. The samples consisted of formalin-fixed, paraffin-embedded (FFPE) colon cancer tissues, collected from Asan Medical Center. Ethical approval was granted by the Institutional Review Board (IRB) with a waiver of informed consent (IRB No. 2021-1698). All procedures conformed to the ethical standards outlined in the Declaration of Helsinki[49]. In this dataset, the voxel values represent the scalar refractive index (RI), and the tissue structures exhibit relatively high spatial complexity and density, making it suitable for evaluating compression under challenging signal conditions.

The Tissue (off-diagonal) dataset contains off-diagonal components of the 3×3 refractive index tensor, representing birefringence information derived from polarization-sensitive HT measurements. These components typically exhibit low signal amplitudes and are more susceptible to noise, providing a stringent testbed for evaluating compression performance on weak, low-contrast signals.

**Filter and compression algorithms for encoding data**

In the encoding pipeline, raw data was first processed using one or more preprocessing filters, followed by compression using a selected algorithm (Fig. 5c). We systematically explored all possible combinations arising from two types of filters and 25 distinct compression configurations.

Two preprocessing filters were evaluated: bit rounding and byte shuffling. These filters enhance compressibility by restructuring the data layout to expose redundancy. The bit rounding filter reduces the numerical precision of floating-point values by setting less significant bits to zero[50]. While this introduces minor information loss, it significantly reduces entropy, improving the effectiveness of downstream compression. Despite its lossy nature, bit rounding is compatible with traditional lossless compression algorithms and is particularly suitable for scientific data where ultra-fine precision is often unnecessary. The byte shuffling filter reorganizes bytes within the data so that similar byte positions across adjacent values are grouped together. This transformation increases local redundancy and improves pattern detection by compression algorithms, thereby boosting compression efficiency without loss of precision.

Following the filtering step, datasets were compressed using 25 configurations encompassing both classic and speed-enhanced compression algorithms (Table 3). The classic algorithms—gzip, LZMA, bzip2, and zlib—rely on dictionary-based encoding and entropy coding schemes[32-35]. These algorithms are well-established in general-purpose data compression and serve as a useful benchmark for evaluating newer alternatives.

| Type | classic | | | | speed-enhanced | | | | |
|---|---|---|---|---|---|---|---|---|---|
| Compressor Name | gzip | LZMA | bzip2 | zlib | LZ4 | zstd | blosc | | |
| | | | | | | | zlib | LZ4 | zstd |
| Levels (L)ow | 1 | 1 | 1 | 1 | . | 1 | 1 | 1 | 1 |
| Levels (M)edium | 5 | 5 | 5 | 5 | . | 11 | 5 | 5 | 5 |
| Levels (H)igh | 9 | 9 | 9 | 9 | 1 | 22 | 9 | 9 | 9 |

Table 3 | List of compression algorithms and levels used in benchmarking.

In contrast, speed-enhanced algorithms were selected for their fast processing speeds and compatibility with parallel workflows. Zstandard (zstd), in particular, is designed to provide high compression ratios while maintaining fast throughput, making it a strong candidate for real-time or cloud-based workflows[37]. We also evaluated the Blosc meta-compression framework, widely used in scientific computing, which supports multithreading and filters such as byte shuffling. Blosc was tested with three backend compressors: zlib, LZ4, and zstd[38].

To assess how varying compression aggressiveness impacts performance, each algorithm was tested at three representative compression levels. The low level prioritized speed and minimal overhead, medium level aimed to balance speed with reasonable data reduction, and the high level sought maximum compression ratio at the expense of processing time. For most algorithms, these correspond to levels 1 (L), 5 (M), and 9 (H). For zstd, which supports a wider compression range, levels 1, 11, and 22 were used. LZ4, designed for ultra-fast compression, was evaluated at its single high-speed setting (level 1), which remains faster than other algorithms across comparable configurations.

This comprehensive setup allowed for a detailed analysis of how different encoding strategies interact with filter combinations and system bandwidth, providing a robust framework for optimizing compression in holotomographic imaging workflows.

## Competing Interests

Prof. YongKeun Park holds financial interests in Tomocube Inc., a company that develops and commercializes holotomography instruments and partially funded this study.

## Acknowledgments

This work was supported by National Research Foundation of Korea grant funded by the Korea government (MSIT) (RS-2024-00442348, 2022M3H4A1A02074314), Institute of Information & communications Technology Planning & Evaluation (IITP; 2021-0-00272) grant funded by the Korea government, Korea Institute for Advancement of Technology (KIAT) through the International Cooperative R&D program (P0028463), and the Korean Fund for Regenerative Medicine (KFRM) grant funded by the Korea government (the Ministry of Science and ICT and the Ministry of Health & Welfare) (21A0101L1-12).

omics data in life sciences. *BMC bioinformatics* **23**, 61 (2022).